\documentstyle[preprint,aps]{revtex}
\makeatletter
\outer\def\newclass #1#2%
    {%
    \edef \ObjectClassList {#1 \ObjectClassList}%
    \expandname\NewCount {#1counter}\global\csname #1counter\endcsname = 0
    \expandname\def     {#1def}{\SaveContents {#1}}%
    \expandname\def    {#1list}{\ListObjects {#1}}%
    \expandname\def  {#1num}##1{\LookUp {#1}##1 \Using {#1counter}#2\Label}%
    }%
\def\NewCount #1
    {%
    \global\advance \count10 by 1
    \global\countdef #1=\count10
    \wlog {\string#1=\string\count\the\count10}%
    }%
\def \expandname  #1#2{\expandafter #1\csname #2\endcsname}%
\def \ifundefined   #1{\expandname \ifx {#1}\relax}%
\let \ObjectClassList = \empty
\newclass {ref}\relax
\newcount \lastrefno	\lastrefno=-1
\newcount \refsequence	\refsequence=0
\def\Refnums#1%
    {%
    \RefRange#1-\end%
    }%
\def\lgycite#1
    {%
    [\RefRange#1-\end]%
    }%
\def\RefRange #1-#2\end%
    {%
    \RefNums #1,\end
    \def \temp {#2}%
    \ifx \temp\empty \else -\expandafter\RefRange \temp\end \fi
    }%
\def\RefNums #1,#2\end%
    {%
    \def \temp {#1}%
    \ifx \temp\empty \else \SkipSpace \temp#1\end\fi
    \ifx \temp\empty
	\ifcase \refsequence
	    \or\or ,\number\lastrefno
	    \else  -\number\lastrefno
	\fi
	\lastrefno = -1
	\refsequence = 0
    \else
	\edef\temp {{ref}\temp\space}%
	\expandafter \LookUp \temp \Using {refcounter}\relax
	\global\advance \lastrefno by 1
	\edef \temp {\number\lastrefno}%
	\ifx \Label\temp
	    \global\advance\refsequence by 1
	\else
	    \global\advance\lastrefno by -1
	    \ifcase \refsequence
		\or ,%
		\or ,\number\lastrefno,%
	    \else   -\number\lastrefno,%
	    \fi
	    \Label
	    \refsequence = 1
	    \ifx\Suffix\empty
		\lastrefno = \Label
	    \else
		\lastrefno = -1
	    \fi
	\fi
	\RefNums #2,\end
    \fi
    }%
\newif\ifSaveFile
\newif\ifnotskip
\newwrite\SaveFile
\def\savefilename {\jobname.sst}%
\def\OpenSaveFile   {\immediate\openout\SaveFile = \savefilename
		     \global\SaveFiletrue}%
\def\CloseSaveFile  {\immediate\closeout\SaveFile
		     \global\SaveFilefalse}%
\def\savestate%
    {%
    \expandafter \SaveCounters \ObjectClassList \end
    }%
\def\SaveCounters #1 #2\end%
    {%
    \def \temp {#1}
    \ifx \temp\empty \else
	\SaveObject {\number\csname #1counter\endcsname}{num}{#1counter}%
	\def \temp {#2}
	\ifx \temp\empty \else \SaveCounters #2\end \fi
    \fi
    }%
\def\Define #1#2#3%
    {%
    \def \temp {#1}
    \ifx \temp\Num	\global \csname #2\endcsname = #3 \else
    \ifx \temp\empty	\expandname\xdef    {#2}{#3}%
    \else		\expandname\xdef {#1_#2}{#3}%
    \fi \fi
    \SaveObject {#3}{#1}{#2}\iftrue
    }%
\def\Contents {\iftrue \SaveContents}%
\def\Num      {num}%
\let\IfDefine     = \Define
\let\IfContents   = \Contents
\def\Savesstout #1#2#3%
    {%
    \immediate\write\@sstout {\noexpand\IfDefine {#2}{#3}{#1}\noexpand\fi}%
    }
\def\SaveObject #1#2#3%
    {%
    \ifSaveFile \else \OpenSaveFile \fi
    \immediate\write\SaveFile {\noexpand\IfDefine {#2}{#3}{#1}\noexpand\fi}%
    \ifnum\SaveFile=\@sstout\else\Savesstout {#1}{#2}{#3}\fi
    }%
\def\SaveContents #1#2%
    {%
    \ifSaveFile \else \OpenSaveFile \fi
    \BreakLine
    \SaveLine {#1}{#2}%
    }%
\begingroup
    \catcode`\^^M=\active %
  \gdef\BreakLine %
    {%
    \begingroup %
    \catcode`\^^M=\active %
    \newlinechar=`\^^M %
    }%
  \gdef\SaveLine #1#2#3%
    {%
    \toks255={#3}%
    \immediate\write\SaveFile %
	{\noexpand\IfContents {#1}{#2}\LBrace\the\toks255\RBrace\noexpand\fi}%
    \endgroup %
    }%
\endgroup
\def\ListObjects #1#2%
    {%
    \ifSaveFile \CloseSaveFile \fi
    \def \ObjClass {#1}
    \let \IfContents = \GetContents
    \let \IfDefine   = \iffalse		\ReadFileList #2,\savefilename,\end
    \let \IfDefine   = \IfDoObject
    \let \IfContents = \iffalse		\input \savefilename
    \let \IfContents = \Contents
    \let \IfDefine   = \Define
    }%
\def\ReadFileList #1,#2\end%
    {%
    \def \temp {#1}%
    \ifx \temp\empty \else \SkipSpace \temp#1\end \fi
    \ifx \temp\empty \else \input #1 \fi
    \def \temp {#2}%
    \ifx \temp\empty \else \ReadFileList #2\end \fi
    }%
\def\GetContents #1#2%
    {%
    \notskipfalse
    \def \temp {#1}
    \ifx \ObjClass\temp \expandname
	\ifx {#1_#2}\relax \else \notskiptrue \fi
    \fi
    \ifnotskip \expandname \DefContents {#1_#2_}%
    }%
\def\DefContents #1#2{\toks255 = {#2} \xdef #1{\the\toks255}}%
\def\IfDoObject #1%
    {%
    \def \temp {#1}
    \ifx \ObjClass\temp \DoObject {#1}
    }%
\def\DoObject #1#2#3%
    {{%
    \expandname\ifx {#1fmt}\relax	\item {#3.}%
    \else				\csname #1fmt\endcsname {#3}\fi
    \expandname\ifx {#1_#2_}\relax	\expandname \CopyLabel {#2}%
    \else				\csname #1_#2_\endcsname \fi
    }}%
\def\CopyLabel #1{\expandafter \Gobble \string #1}

\def\LookUp #1#2 #3\Using#4#5%
    {%
    \expandname \ifx {#1_#2}\relax
	\global\advance \csname #4\endcsname by 1
	\expandname \xdef {#1_#2}{\number \csname #4\endcsname}%
	\let \temp = #5%
	\ifx \temp\relax \else \expandname \temp {#1_#2}\fi
	\expandname \SaveObject {#1_#2}{#1}{#2}%
    \fi
    \xdef \Label {\csname #1_#2\endcsname}%
    \gdef \Suffix {#3}%
    \ifx \Suffix\empty \else
	\xdef \Suffix {\expandafter\TrimSpace \Suffix\end}%
	\xdef \Label  {\Label\Suffix}%
    \fi
    }%

\def\Gobble	 #1{}%
\def\TrimSpace   #1 \end{#1}%
\def\SkipSpace   #1#2#3\end%
    {%
    \def \temp {#2}%
    \ifx \temp\space \SkipSpace #1#3\end
    \else \gdef #1{#2#3}\fi
    }%
\begingroup
\catcode`\<=1 \catcode`\{=12
\catcode`\>=2 \catcode`\}=12
\xdef\LBrace<{>%
\xdef\RBrace<}>%
\endgroup

\NewCount\RefOne	\NewCount\RefTwo
\NewCount\NumKnown	\NewCount\NumUnknown
\NewCount\NumChanged	\NewCount\LastNumChanged
\NewCount\IfOrdered	\NewCount\Done
\newcount\dummy 
\def\cite#1{\NumKnown=0\NumUnknown=0\Done=0\gdef\List{|<>#1,}%
    \loop\expandafter\reFsorT \List,\end\ifcase\Done\repeat%
    \ifcase\NumKnown\else%
        \NumChanged=0\LastNumChanged=0\Done=0%
        \loop\expandafter\sorTcitE \List\end\ifcase\Done\repeat%
    \fi\expandafter\citE \List\end}
\def\citE #1\end{\lgycite{#1}}
\def\reFsorT |#1<#2>#3,#4,\end{\def\temp{#3}%
    \ifx\temp\empty\Done=1
        \ifcase\NumKnown\gdef\List{#2}%
        \else%
	    \ifcase\NumUnknown\gdef\List{|<>#1,|}%
            \else\gdef\List{|<>#1,|#2}%
            \fi%
        \fi%
    \else\expandname \ifx {ref_#3}\relax%
             \ifcase\NumUnknown\gdef\List{|#1<#3>#4,}%
             \else\gdef\List{|#1<#2,#3>#4,}%
             \fi\global\advance\NumUnknown by 1%
         \else%
             \ifcase\NumKnown\gdef\List{|#3<#2>#4,}%
             \else\gdef\List{|#1,#3<#2>#4,}%
             \fi\global\advance\NumKnown by 1%
         \fi%
    \fi}
\def\storeone#1{\global\RefOne=#1}
\def\storetwo#1{\global\RefTwo=#1}
\def\Ordered#1#2{%
    \expandafter\storeone\csname ref_#1\endcsname%
    \expandafter\storetwo\csname ref_#2\endcsname%
    \gdef\temp{\number\RefOne}\gdef\temp{\number\RefTwo}
    \ifnum\RefOne<\RefTwo\IfOrdered=0\else\IfOrdered=1\fi%
    \gdef\temp{\number\IfOrdered}}
\def\sorTcitE |#1<#2>#3,#4|#5\end{%
    \def\tempa{#1}\def\tempb{#2}\def\tempc{#3}\def\tempd{#4}\def\tempe{#5}%
    \ifx\tempb\empty
        \ifx\tempc\empty
            \ifx\tempe\empty%
                \ifnum\NumChanged=\LastNumChanged\Done=1\gdef\List{#1}%
                \else\global\LastNumChanged=\NumChanged\gdef\List{|<>#1,|}%
                \fi%
            \else%
                \ifnum\NumChanged=\LastNumChanged\Done=1\gdef\List{#1,#5}%
                \else\global\LastNumChanged=\NumChanged\gdef\List{|<>#1,|#5}%
                \fi%
            \fi%
        \else
            \gdef\List{|<#3>#4,|#5}%
        \fi%
    \else
        \ifx\tempc\empty
            \ifx\tempa\empty\gdef\List{|#2<#3>#4,|#5}%
            \else\gdef\List{|#1,#2<#3>#4,|#5}%
            \fi%
        \else
            \Ordered{#2}{#3}\ifcase\IfOrdered%
                \ifx\tempa\empty\gdef\List{|#2<#3>#4,|#5}%
                \else\gdef\List{|#1,#2<#3>#4,|#5}%
                \fi%
            \else%
                \global\advance\NumChanged by 1%
                \ifx\tempa\empty\gdef\List{|#3<#2>#4,|#5}%
                \else\gdef\List{|#1,#3<#2>#4,|#5}%
                \fi%
            \fi%
        \fi%
    \fi%
    }
\let\@mainsst=\SaveFile
\let\@sstout=\@mainsst
\def\journal#1,#2,{{\it #1\/} {\bf #2},}
\def\jpc#1,{\journal J. Phys. Chem., #1,}
\def\jcp#1,{\journal J. Chem. Phys., #1,}
\def\jpp#1,{\journal J. Phys. (Paris), #1,}
\def\jpf#1,{\journal J. Phys. France, #1,}
\def\epl#1,{\journal Europhys. Lett., #1,}
\def\prl#1,{\journal Phys. Rev. Lett., #1,}
\def\pr#1,{\journal Phys. Rev., #1,}
\def\macromol#1,{\journal Macromolecules, #1,}

\def\deldot{\vec\nabla\kern -2pt\cdot}
\def\delcross{\vec\nabla\kern -2pt\times}

\def\\{\relax \ifmmode \backslash \else {\tt\char`\\}\fi }

\let\int=\intop         
\def\lsim{\mathrel{\mathpalette\@versim<}}
\def\gsim{\mathrel{\mathpalette\@versim>}}
\def\@versim#1#2{\vcenter{\offinterlineskip
	\ialign{$\m@th#1\hfil##\hfil$\crcr#2\crcr\sim\crcr } }}

\def\undertext #1{\vtop{\hbox{#1}\kern 1pt \hrule}}
\def\references{\section*{References\@mkboth
  {REFERENCES}{REFERENCES}}\list
  {[\arabic{enumi}]}{\settowidth\labelwidth{[99]}\leftmargin\labelwidth
    \advance\leftmargin\labelsep\itemsep=0pt\parsep=0pt
    \usecounter{enumi}}
    \def\newblock{\hskip .11em plus .33em minus .07em}
    \sloppy\clubpenalty4000\widowpenalty4000
    \sfcode`\.=1000\relax}

\makeatother
\def\mob{{\cal M}}
\begin{document}
\preprint{NSF ITP 94-117}
\tighten
\title{Swelling Kinetics of Layered Structures:\\ Triblock Copolymer Mesogels}
\author{C.-M. Chen$^{1,2}$, F.C. MacKintosh$^{1,2}$,
and D.R.M. Williams$^{1,2,3}$}

\address{$^1$Department of Physics, University of Michigan, Ann Arbor,
MI 48109-1120.\\ $^2$Institute for Theoretical Physics,
University of California at Santa Barbara, CA 93106-4030.\\
$^3$ Institute of Advanced Studies, Research School of
Physical Sciences and Engineering, \\  The Australian National University,
Canberra.}
\maketitle

\begin{abstract}
We consider the swelling kinetics of layered structures. We focus on the
case of triblock copolymer mesogels, although our results are
applicable to other layered structures including clays.
We assume the mesogels are swollen by a solvent that is good for the
bridging block but poor for the non-bridging block.
At long times the penetration front moves as in ordinary diffusion,
i.e., as $t^{1/2}$.
At short times, however, the bending elasticity of the non-bridging
layers becomes important.
This bending elasticity leads to a
$t^{1/6}$ relaxation of the penetration front at early times.
The crossover length between these two regimes is approximately the
width of a single layer.
However, for a large number of lamellae there is a cooperative
effect which leads to a large enhancement of this crossover length.
\end{abstract}
\newpage
\section{Introduction} In the strong segregation limit, a symmetric
diblock copolymer with $A$ and $B$ blocks will form a lamellar phase
consisting of a series of $AB$ layers.  This kind of system has
received much study and may have applications as a novel composite
material.  In general, however, it suffers from one major drawback:  the
AB lamellae are only weakly bonded to one another by local Van der
Waals forces, and possibly by a few entanglements.  The system thus
lacks mechanical integrity in the melt state and the $AB$ interfaces
are very weak.  This is the case even if both blocks are glassy.  If
for instance the system is placed in a solvent which is good for the
$B$ blocks, these blocks swell and the individual lamellae separate
from one another.  However, as noted by Halperin and Zhulina
\cite{Meso}, one very simple modification allows much greater
mechanical integrity.  This is to use an $ABA$ triblock copolymer.
A certain fraction of the $B$ chains bridge between layers.
In the melt
state this system, though a liquid, is much more difficult to pull
apart, because of the presence of these bridges.
If the $A$ regions
are rubbery or glassy then the system becomes a solid, even though the
$B$ regions may still show liquid-like behavior.  When the $B$ regions
are swollen by a selective solvent (which is good for $B$ and
bad for $A$), the
system becomes a sandwich with rubbery liquid layers between solid
regions.  This swollen system was given the name ``mesogel"
\cite{Meso}.  Although the theoretical studies of these systems
have been rather recent, mesogels have existed in practice
for much longer \cite{Oldmeso}.  Their mechanical integrity offers
many advantages over diblock systems, and they are already being used
in controlled drug release systems \cite{release}.

Previous theoretical studies have concentrated on the equilibrium behaviour
and to a lesser extent on the rheology of mesogels \cite{Meso}.
The questions of interest for equilibrium properties are the
degree of bridging,
the  degree of swelling, and the mechanical moduli of the composite
system. Here our interest lies in the kinetics of swelling, i.e.,
the approach to equilibrium. The swelling kinetics of isotropic gels
is a subject for which there is a broad literature reporting many novel
effects \cite{Kinetics,Tanaka,Sekimoto}. Mesogels, are by definition
non-isotropic, and might exhibit interesting swelling behaviour
because of this as well as the competition  between layer
elasticity and solvent penetration. The swelling of lamellar mesogels
is related to  the swelling of other layered systems such as clays
\cite{Clay}. These have numerous applications, particularly
in environmental science.
Geometrically the two systems are similar, although there are major
differences in both the type and magnitude of the physical forces
involved. In clays, for instance, the layers are much smaller and
electrostatic and
short-range forces are thought to be of great importance. Another
related area is the swelling of polymeric glasses. In that case
there are two kinds of temporal behaviour, ordinary diffusion and
``diffusion" linear in time. There have been many proposals for
why this occurs \cite{TW,Cohen}, all of which agree more-or-less
with the experimental results.  However, at present there is no widely
accepted definitive theory, although elastic effects almost certainly play
some role.

In any system involving mesogels the gels must at some time go
through a swelling process, and our study is applicable to all
such cases. The opposite case where one de-swells a mesogel may be
important in drug delivery systems \cite{rel2}. The same model
can be used to describe de-swelling, although we do not discuss it
further here.
In section II we examine the case of a single layer of a mesogel,
such as might be obtained by spin-coating. This turns out to have a
mesoscopic length $L_1$, below which the elasticity of the layers
needs to be accounted for, and for which the
swelling kinetics are non-diffusive in character.
It turns out that $L_1$ is of order the layer thickness.
In section III we generalise this to more macroscopic multilamellar samples.
There, the crossover length becomes much larger: $L_M \simeq M^{1/2} L_1$,
where $M$ is the number of layers. Hence $L_M$ can become of
macroscopic size. We conclude with a discussion in section IV.

\section{The Model of diffusion process in single layer}
Consider a melt of $ABA$ triblocks, with blocks of roughly equal
length. At low temperatures this system will microphase separate
to form a series of distinct $B$ layers separated by $A$ layers,
with sharp boundaries in between. In general this will have a large
number of defects, which can be removed by shearing.
At lower temperatures, or if
the $A$ phase is crosslinked, the $A$ region may become a
rubbery solid. The system then consists of a series of rubbery $A$
layers separated by  molten  $B$ layers. This results in a well-ordered
``meso-rubber". This system can be placed in a selective
solvent that is good
for the $B$ block and poor for the $A$ block. We assume for simplicity
that only one edge of the specimen is in contact with the solvent reservoir.
The solvent penetrates the $B$ regions and swells them. Here we
consider a single $ABA$ lamella. Initial contact of the solvent with
the $B$ regions is energeticly favourable but is opposed by two
effects. The first is the  stretching of the $B$ chains to form a
brush. The second is the bending and stretching of the surrounding
$A$ regions. This implies that the free energy per unit area of the
system consists of three terms
\begin{eqnarray}
F= F_{B~\rm stretch} + F_{B~\rm interaction} + F_{A~\rm bend}.
\label{all}
\end{eqnarray}
Here we ignore the stretching energy of $A$ domains, which is
asecond-order effect in relaxed layers.  We need to evaluate each of these
terms in the partially swollen state to describe the swelling
kinetics.

Before swelling, the $B$ regions form a dense melt of thickness
$\bar{h}$.  After equilibrium swelling is reached, the $B$ regions form a
swollen polymer brush \cite{Rev,Rev2} with the same grafting density as the
melt.  In
the swollen state, dense grafting enforces strong overlap among the
undeformed coils.  For tethered chains in a good solvent, this
increases the number of monomer-monomer contacts and the corresponding
interaction energy.  This penalty is reduced by stretching the chains
along the normal to the grafting sites, thereby lowering the monomer
concentration in the layer and increasing the layer thickness $h$.
Stretching lowers the interaction energy per chain,
$F_{B~\rm interaction}$, at the price of a higher elastic free energy,
$F_{B~\rm stretch}$.  The interplay of these two terms sets the
equilibrium thickness of the layer $h_{eq}$.

We write the layer thickness of the $B$ region in the partially
swollen state as
\begin{eqnarray}
h\left( x \right) = \bar{h} \left( 1+\psi \left( x \right) \right),
\label{height}
\end{eqnarray}
where $\psi$ is the solvent concentration in the $B$ region.  We can
express the volume fraction of solvent $\phi_s$ and the volume
fraction of polymer $\phi_p$ as $\phi_s=\psi/(1+\psi)$ and $\phi_p =
1/(1+ \psi)$.  In the Alexander-de Gennes model, the $B$ layer is
envisioned as a close-packed array of blobs of uniform size $\xi$, which
is equal to the grafting distance $d$.
Each of these blobs costs a free energy of
$kT$.  The free energy of the $B$ layer per unit area is given by \cite{Rev}
\medskip
\begin{eqnarray}
F_B & = & F_{B~\rm interaction} + F_{B~\rm stretch} \nonumber \\
    & = &  kT \left( h \over a^3 \right) \phi_p^{9 \over 4}+
      kT \left( h^2 \over {N{a^2} {d^2}} \right)\phi_p^{1 \over 4},
\label{Alex1}
\end{eqnarray}
\medskip
where numerical prefactors have been ignored,
and $a$ is the $B$ monomer size.
Since $\phi_p=(1+\psi)^{-1}$ and $h=\bar{h} (1+\psi)$,
Eq.\ (\ref{Alex1}) can be rewritten as
\medskip
\begin{eqnarray}
F_B & = &  kT \left( \bar{h} \over a^3 \right) (1+\psi)^{-{5 \over 4}}
+ kT \left( \bar{h}^2 \over {N{a^2} {d^2}} \right)(1+\psi)^{7\over 4}.
\label{Alex2}
\end{eqnarray}
\medskip
Minimising this yields the equilibrium swelling
$\psi_{eq}={\left( {5 Nd^2} / {7 a \bar{h}} \right)}^{1 \over 3}-1$.
It is convenient to simplify $F_B$ by expanding about $\psi_{eq}$,
to yield an approximate free energy per unit area for the $B$ regions
\cite{qu,Mi}
\medskip
\begin{eqnarray}
F_B \simeq  { 1 \over 2} S (\psi- \psi_{eq})^2.
\label{expand}
\end{eqnarray}
\medskip
Here, the coefficient $S$ in Eq.\ (\ref{expand}) is given by
\medskip
\begin{eqnarray}
S & = & {{\partial^2} F_B \over \partial \psi^2}  \nonumber \\
  & = &  kT \left( {\bar{h}^2 \over Na^2 d^2} \right)
        \left( {\bar{h} a \over Nd^2} \right)^{1 \over 12}.
\label{s}
\end{eqnarray}
\medskip

The remaining term in the free energy is the bending elastic energy
of the $A$ regions:
\medskip
\begin{eqnarray}
F_A = {1\over2} \kappa \left({\mbox{\boldmath{$\nabla$}}}^2 h \right)^2
= {1\over2} \kappa^{\prime} \left({\mbox{\boldmath{$\nabla$}}}^2 \psi
\right)^2,
\label{bending}
\end{eqnarray}
\medskip
where $\kappa$ is the bending elastic constant and $\kappa^{\prime} =
\bar{h}^2 \kappa$.  If the $A$ regions form a rubbery solid, then we can
estimate $\kappa$ from elasticity theory as $ \kappa \sim \mu_r H_A^3$
where $\mu_r$ is the modulus of rigidity
of the $A$ regions and $H_A$ is the
thickness of the $A$ layers \cite{LandauLifshitz}.
This modulus is related to the number of
crosslinks per unit volume $n$, by $\mu_r \sim n kT$.

The chemical potential of the solvent is
\begin{eqnarray}
\mu & = & {\delta F} \over {\delta \psi}  \nonumber \\
    & = & \kappa^{\prime} \left({\mbox{\boldmath{$\nabla$}}}^4 \psi \right)+
          S \left( \psi - \psi_{eq} \right).
\label{chepot}
\end{eqnarray}
The solvent current is defined by
${\bf J}_s = - {\mob} {\mbox{\boldmath{$\nabla$}}} \mu$, where
${\mob}$ is a mobility, which we take to be independent of the
concentration $\psi$ \cite{Broch}.
{}From the continuity equation
\begin{eqnarray}
{{\partial \psi} \over {\partial t}} + {\mbox{\boldmath{$\nabla$}}}
\cdot {\bf J}_s =0,
\label{continuity}
\end{eqnarray}
we have the diffusion equation
\begin{eqnarray}
{\partial \psi} \over {\partial t}
& = & {\mbox{\boldmath{$\nabla$}}} \cdot \left( {\mob}
{\mbox{\boldmath{$\nabla$}}}
      \mu \right)  \nonumber \\
& = & \kappa^{\prime} {\mob} \left( {\mbox{\boldmath{$\nabla$}}}^6  \psi
\right)
      +S {\mob} \left( {\mbox{\boldmath{$\nabla$}}}^2  \psi \right).
\label{diff}
\end{eqnarray}
Here we consider a single layer parallel to the $xy$ plane,
with infinite extent in the $y$ direction (Fig.\ \ref{one}). The solvent
lies in the region $x<0$ and the gel in the region $x>0$.
This reduces the problem to one dimension, with spatial gradients
only in the $x$ direction.

{}From Eq.(\ref{diff}), we expect the diffusion process to be dominated by
the bending energy at shorter lengthscales (or earlier times), and to be
dominated by the interaction energy at longer lengthscales (or later
times).
Therefore, the diffusion process is characterized by a
crossover time scale $T_1$ and a crossover length $L_1$.
Depending on the initial conditions, for times less than $T_1$
or for solvent penetration less than $L_1$, the process is not characterized
by simple diffusion.
Instead, the profile spreads as $t^{1/6}$.
At later times, as the solvent profile spreads out,
simple diffusion dominates.
In most of what follows, we shall be concerned primarily with
the non-diffusive regime, and in particular with crossover length,
below which it can be observed.
We can estimate the crossover length $L_1$ by the substitution
${\mbox{\boldmath{$\nabla$}}} \rightarrow 1/L_1$.  From Eq.\
(\ref{diff}), the crossover length is approximately
\medskip
\begin{eqnarray}
L_1 \simeq \left( {\bar{h}^2 \kappa} \over S \right)^{1 \over 4}.
\label{L1}
\end{eqnarray}
\medskip
In a rubbery solid where $\kappa \simeq kT n H_A^3$, the crossover length
$L_1$ becomes
\medskip
\begin{eqnarray}
L_1 \simeq \left[ n H_A^3 Na^2 d^2
    \left({d \over a}\right)^{1 \over 3} \right]^{1 \over 4}.
\label{L2}
\end{eqnarray}
\medskip
Within the Alexander-de Gennes approximation, the chains are assumed to be
equally stretched.  The grafting distance is estimated to be
\medskip
\begin{eqnarray}
d \simeq\left({4 kT \over \gamma a^2} \right)^{1 \over 6} N^{1 \over 6}a.
\label{d}
\end{eqnarray}
\medskip
We can also estimate the layer thickness of $A$ domains as
\begin{eqnarray}
H_A & = & N a^3 \over d^2  \nonumber \\
  & \simeq & \left({4 kT \over \gamma a^2} \right)^{-{1 \over 3}}
        N^{2 \over 3}a.
\label{zz}
\end{eqnarray}
\medskip
The layer thickness of the $B$ domains, prior to swelling is similarly
given by
$\bar{h} \simeq a N^{2/3} $.
Substitution of Eqs.\ (\ref{d}) and (\ref{zz}) into Eq.\ (\ref{L2}) leads
to a crossover length
\medskip
\begin{eqnarray}
L_1 \simeq (n a^3)^{1/4} N^{13/72} \bar{h}
\label{L3}
\end{eqnarray}
\medskip
If we consider a tightly crosslinked network so that $na^3\sim 1$,
then the crossover length is roughly the layer thickness $\bar{h}$.
This is a mesoscopic length of order $10^2 $ to $10^3$ Angstroms.
Note, however, that if the modulus of the $A$ regions is very large
(say, for instance, if the $A$ regions were glassy),
then the crossover length $L_1$
could be much larger. Before the solvent has reached $x=L_1$, we expect
the solvent front moves as $t^{1/6}$. At later times it obeys ordinary
diffusion---moving as $t^{1/2}$. The anomalous $t^{1/6}$ ``diffusion" is a
result of the bending elasticity of the $A$ layers.
A similar anomalous exponent can be found in the hydrodynamics of
membrane systems \cite{BL,Rami}.

\section{The Model of diffusion process in multilayers}
In the above we considered the swelling of a single layer.
In that case, the $A$ layer elasticity only affected the initial
swelling up to a length scale of order of the layer thickness.
For a multilayer system, this effect can be much more pronounced.
As illustrated in Fig.\ \ref{two}, the swelling of
the inner layers is transmitted to the outer layers,
and hence the outer layers must bend
significantly more. In order to examine this, we first consider
the case of uniform swelling, in which each layer is swollen
by the same amount at a given value of $x$. Below, we shall relax
this assumption. For $M$ layers, uniform swelling corresponds to
$\psi_m (x)=\psi(x)$ for $m=1, 2, \cdots M$.
The height of $m$-th layer is given by
\medskip
\begin{eqnarray}
h_m \left( x \right) = m \bar{h} \left( 1+\psi \left( x \right) \right).
\label{hm}
\end{eqnarray}
\medskip
The free energy of the mesogel is
\medskip
\begin{eqnarray}
F & = & \sum_{m=1}^{M}\left[ {1\over2} \kappa
\left(\nabla^2 h_m \right)^2
+{1 \over 2}S \left( \psi_m - \psi_{eq} \right)^2 \right],\nonumber \\
  & \simeq & {M^3  \over 6} \kappa^{\prime}
  \left(\nabla^2 \psi \right)^2+
  {S M \over 2} \left( \psi - \psi_{eq} \right)^2.
\label{Funi}
\end{eqnarray}
\medskip
Thus
\medskip
\begin{eqnarray}
{\partial \psi \over \partial t} \simeq \kappa^{\prime} {\mob}
{M^3 \over 3} \nabla^{6} \psi + S {\mob} M \nabla^{2} \psi.
\label{chemuni}
\end{eqnarray}
\medskip
In other words, the relaxation rate for a mode of wavevector
$q$ in the $x$ direction, is
\medskip
\begin{eqnarray}
{\tau_q}^{-1} \simeq {\mob} \left( \kappa^{\prime}{M^3 \over 3}q^6+
S M q^2 \right).
\label{rate}
\end{eqnarray}
\medskip
The characteristic length $L_M$, beyond which the swelling becomes
dominated by ordinary diffusion, is found by equating the two terms
on the right hand side of (\ref{rate}) with $q = 1/L_M$, and is
\begin{eqnarray}
L_M  \approx  M^{1\over 2} L_1,
\label{L4}
\end{eqnarray}
where $L_1$ is the crossover length for a single layer.
For a large number, $M$, of layers,
this is a lengthscale much larger than the thickness of a single
lamella.
However, since we consider the case of
swelling that occurs from an edge, our results are valid
for a mesoscopic crossover length $L_M$ less than the horizontal
dimensions of the sample in Fig.\ 2.
The corresponding crossover time is
\begin{equation}
\tau_M\sim M{\bar h}^2/D,
\end{equation}
where $D$ is the diffusion constant for the solvent.

To go beyond the approximation of uniform swelling, we express
the height of the m-th layer in a more general form
\begin{eqnarray}
h_m \left( x \right) - h_{m-1} \left( x \right)
= \bar{h} \left( 1-\psi_m \left( x \right) \right),
\label{mheight}
\end{eqnarray}
where $m = 1, 2, \cdots M$, and $h_0 \equiv 0$.
The free energy of the mesogel is
\medskip
\begin{eqnarray}
F & = & {1 \over 2} \sum_{m=1}^{M}\left\{ \kappa
\left( \nabla^2 h_m \right)^2
+ {S \over {\bar{h}}^2}
\left[ h_m-h_{m-1}-\bar{h}(1+\psi_{eq}) \right]^2 \right\}.
\label{Fg}
\end{eqnarray}
\medskip
The chemical potential of $l$-th layer can be calculated as
\medskip
\begin{eqnarray}
\mu_l & = & \sum_{m=1}^{M} {\delta F \over \delta h_m \left( x \right)}
            {\partial h_m\left( x \right) \over
            \partial \psi_l\left( x \right)} \nonumber \\
      & = & \kappa \bar{h} \sum_{m=l}^{M} \nabla^4 h_m
            +{S \over {\bar{h}}} \sum_{m=l}^{M-1}\left( 2 h_m - h_{m+1}
            - h_{m-1}\right)
            +{S \over \bar{h}}\left[h_M-h_{M-1}-\bar{h}(1+\psi_{eq})\right].
\label{mug}
\end{eqnarray}
\medskip
The dynamical equations for $h_l$ are given by
\begin{eqnarray}
{{\partial (h_l - h_{l-1})} \over {\partial t}}
 = {\mob} \bar{h}^2 \kappa \sum_{m=l}^{M} \nabla^6 h_m
   &+&{\mob} S \sum_{m=l}^{M-1} \nabla^2 \left( 2 h_m-h_{m+1}-h_{m-1}\right)
\nonumber \\
            &+&{{\mob}S}\nabla^2 \left( h_M - h_{M-1} \right),
\label{diff2l}
\end{eqnarray}
and
\begin{eqnarray}
{{\partial (h_M - h_{M-1})} \over {\partial t}}
  = \kappa {\mob} \bar{h}^2 \nabla^6 h_m
   +{\mob} S \nabla^2 \left(  h_M - h_{M-1} \right).
\label{diff2M}
\end{eqnarray}

For large $M$ and small displacements of the layer heights
away from their initial positions,
we characterize the layer heights by a displacement field
$u \left( x,z \right)$:
\begin{eqnarray}
h_m \left( x \right) =m \bar{h} + u \left( x,m\bar{h} \right),
\label{displacement}
\end{eqnarray}
Eq.\ (\ref{diff2l}) becomes
\begin{eqnarray}
{\partial \over \partial t}
\left( {\partial^2 u \over \partial z^2} \right)
= - {\mob} \kappa \nabla^6 u \left( x,z \right) + {\mob} S \nabla^2
\left( {\partial^2 u \over \partial z^2} \right)
\label{gendiff}
\end{eqnarray}
after differentiating once with respect to $z$.
(A discrete expression of this can
also be obtained by subtraction of Eq.\ [\ref{diff2l}] for
adjacent $l$.)
If we express the displacement field $u(x, z)$
as a sum over wavevectors in the $x$ and $z$ directions, $q$ and $k_z$,
\begin{eqnarray}
u \left( x,z \right) = \sum_{q,k_z} \tilde{u} \left(q, k_z \right)
e^{i \left( q x +k_z z \right)},
\label{transform}
\end{eqnarray}
then the  relaxation rate as a function of $q$ and $k_z$ is
\begin{eqnarray}
\tau_q^{-1} = {\mob} {\kappa } q^6 k_z^{-2} + {\mob} {S } q^2.
\label{mrate}
\end{eqnarray}
The initial swelling mode corresponds to the smallest
value of $k_z$ consistent with the boundary conditions, namely
\begin{eqnarray}
k_z \approx {1 \over M \bar{h}}.
\label{fastmode}
\end{eqnarray}
Thus, the crossover length is
\begin{eqnarray}
L_M & \approx & \left( {\kappa \bar{h}^2 \over S} M^2 \right)^{1 \over 4}
\nonumber \\
    & \approx & M^{1 \over 2} L_1.
\label{L5}
\end{eqnarray}
i.e. the same result as was found in the uniform swelling case (\ref{L4}).

Equation (\ref{Fg}) can also be expressed in terms of the
concentration variables $\psi_m(x)$:
\medskip
\begin{eqnarray}
F & = & {1 \over 2} \sum_{m=1}^{M}\left\{ \kappa
\left( \nabla^2 h_m \right)^2
+ {S}
\left( \psi_m-\psi_{eq} \right)^2 \right\},
\label{Fgpsi}
\end{eqnarray}
\medskip
where
\medskip
\begin{eqnarray}
h_m = \bar{h} \sum_{l=1}^{m} \left( 1 + \psi_l \left( x \right) \right).
\label{sumcon}
\end{eqnarray}
\medskip
The chemical potential of $l$-th layer is given by
\medskip
\begin{equation}
\mu_l  =  \kappa \bar{h}^2 \sum_{m=l}^{M} \sum_{m'=1}^{m} \nabla^4 \psi_{m'}
            +{S } \left(\psi_l-\psi_{eq}\right).
\label{mug2}
\end{equation}
\medskip
The dynamical equations for $\psi_l$ can be expressed as
\begin{equation}
\left(\begin{array}{c}
         {\partial\over\partial t}\psi_1 \\
         \vdots \\
         {\partial\over\partial t}\psi_{M-1} \\
         {\partial\over\partial t}\psi_M
      \end{array} \right)
=\kappa \bar h^2 {\mob}
\left(\begin{array}{cccc}
         M & \cdots & 2 & 1 \\
         \vdots &   & \vdots & \vdots \\
         2 & \cdots & 2 & 1 \\
         1 & \cdots & 1 & 1
      \end{array} \right)
\left(\begin{array}{c}
         \nabla^6\psi_1 \\
         \vdots \\
         \nabla^6\psi_{M-1} \\
         \nabla^6\psi_M
      \end{array} \right)
+S{\mob}
\left(\begin{array}{c}
         \nabla^2\psi_1 \\
         \vdots \\
         \nabla^2\psi_{M-1} \\
         \nabla^2\psi_M
      \end{array} \right).
\end{equation}
At early times, for which  the spatial gradients are large, the
dominant relaxation rate corresponding to a mode of wavevector
$q$ is
\begin{equation}
\tau_q^{-1}=\kappa \bar h^2{\mob} \lambda^{\rm (max)}q^6+S {\mob} q^2,
\end{equation}
where $\lambda^{\rm (max)}$ is the largest eigenvalue of the matrix
\begin{equation}
\left(\begin{array}{cccc}
         M & \cdots & 2 & 1 \\
         \vdots &   & \vdots & \vdots \\
         2 & \cdots & 2 & 1 \\
         1 & \cdots & 1 & 1
      \end{array} \right).
\end{equation}
It can be shown that the eigenvalues of this matrix are given by \cite{Guitter}
\equation
\lambda={1\over2\left(1-\cos\omega\right)},
\endequation
where $\omega$ satisfies
\equation
\tan\left(M\omega\right)=\cot{\omega\over2}.
\endequation
Thus, for large $M$, the leading eigenvalue is
\equation
\lambda^{\rm (max)}\simeq{4\over\pi^2}M^2\left(1+{1\over M}\right).
\label{lammax}
\endequation
(This was also determined by numerical evaluation of
$\lambda^{\rm (max)}$ for $M$ up to $100$.)
Thus, the crossover length
\begin{equation}
L_M\simeq L_1M^{1/2}\left(1+{1\over4M}\right).
\label{LM}
\end{equation}

On simple physical grounds it is expected that the anomalous
swelling at early times will be most apparent for the inner layers,
since the curvature elasticity of the
outer layers tends to constrain the inner layers.
In other words, the initial relaxation of a profile such as the one
shown in Fig.\ 2 will be dominated by a mode that relaxes gradients
of the inner layers ($h_1(x), h_2(x), \ldots$) faster than
the outer layers ($\ldots, h_{M-1}(x), h_M(x)$).
This is reflected in the eigenvector corresponding to this mode
(Eq.\ [\ref{lammax}]), which is given by
\equation
\psi^{\rm (max)}_l\propto
\sin\left({(M-l+1)\pi\over 2M}\right).
\endequation
Thus, indeed the initial relaxation of the
inner layers (small $l$) is greater than that of the outer layers (large
$l$).
However, the crossover length in Eq.\ (\ref{LM}) still grows as $M^{1/2}$,
as for uniform swelling.

\section{Discussion}
In this paper we have examined the swelling kinetics of lamellar
mesogels. For such mesogels the elasticity of the solid lamellae
leads to a novel kinetics which is non-diffusive at short times.
Similar effects should be important for the swelling of clays and
other layered structures.
There is one better-known case of non-diffusive behaviour,
and that occurs in the swelling of isotropic polymer glasses
\cite{TW,Cohen}. In that system there occurs both case I behaviour,
which obeys ordinary diffusion, and case II behaviour
where the front moves linearly in time. There is even a
``super-case II" behaviour which has a front position moving
as $t^{3/2}$. In general the explanations of case II involve
either a diffusion constant which strongly depends on concentration,
or a stress relation which similarly depends strongly on the
concentration. Lamellar mesogels represent a totally different
type of behaviour, where the front moves as $t^{1/6}$, at early times
and as $t^{1/2}$ at later times. Note that here we have
ignored the dependence of the diffusion constant on concentration
and have expanded the free energy about the equilibrium state.
Our results are thus most applicable to the situation where
the meso-rubber is first pre-swollen and then swollen again to
reach an equilibrium swelling. However, the general effect of
the $A$ layer elasticity will always be to modify the swelling
at early times.  We have also ignored cracks in our analysis.
If the strains set up in the system are very large then cracks
may appear in the $A$ region. Such cracks are indeed seen
in the swelling of polymer glasses, which can swell explosively.

\acknowledgements The authors wish to thank E. Guitter for useful
discussions. D.R.M.W. thanks P.G. de Gennes
for useful discussions on case II swelling.
The authors
acknowledge partial support from the Donors of the Petroleum Research
Fund, administered by the American Chemical Society, from the Exxon
Education Fund, and from the NSF under Grant Nos. DMR-91-17249,
DMR-92-57544 and PHY-89-04035.
D.R.M.W also acknowledges the support of a QEII research fellowship.

\let\j=\journal
\refdef{Guitter}{Guitter, E. private communication.}
\refdef{Oldmeso}{Legge, N.R.; Holden, G.; Schroder, H.E., Eds.;
{\it Thermoplastic Elastomers}; Hanser Publishers: Munich, 1987.
Keller, A.; Odell, J.A. In {\it Processing, Structure and Properties of
Block Copolymers}; Folkes, M.J., Ed.; Elsevier: New York, 1985.}
\refdef{Rami}{Marathe, Y.; Ramiswamy, S.
\j Europhys. Lett., 1989, {\it 8}, 581.}
\refdef{BL}{Brochard F.; Lennon, M.
\j J. Phys. (Paris), 1975, {\it 36}, 1035.}
\refdef{Rev}{Halperin, A.; Tirrell, M.; Lodge, T.P.
\j Adv. Polym. Sci., 1992, {\it 33}, 100.}
\refdef{Rev2}{Milner, S.T.
\j Science, 1991, {\it 251}, 905.}
\refdef{TW}{Thomas, N.L.; Windle, A.H.
\j Polymer, 1982, {\it 23}, 529.}
\refdef{Cohen}{Hayes, C.K.; Cohen, D.S.
\j J. of Polym. Sci. B, 1992, {\it 30}, 145 (and references therein).}

\refdef{Broch}{Brochard, F.; de Gennes, P.G.
\j Physicochem. Hydrodyn., 1983, {\it 4}, 313.}
\refdef{Meso}{Zhulina, E.B.
\j Macromolecules, 1993, {\it 26}, 6273.
Misra, S.; Varanasi, S.
\j Macromolecules, 1993, {\it 26}, 4184.
Eicke, H.F.; Gauthier, M.; Hammerich, H.
\j J. Phys. II, 1993, {\it 3}, 255.
Zolzer, U.; Eicke, H.F.
\j J. Phys. II, 1992, {\it 2}, 2207.
Zhulina, E.B.; Halperin, A.
\j Macromolecules, 1992, {\it 25}, 5730.
Halperin, A.; Zhulina, E.B.
\j Europhys. Lett., 1991, {\it 16}, 337.}

\refdef{Clay}{Greathouse, J.A.; Feller, S.E.; Mcquarrie, D.A.
\j Langmuir, 1994, {\it 10}, 2125.
Huerta, M.M.; Curry, J.E.; Mcquarrie, D.A.
\j Clay Clay M., 1992, {\it 40}, 491.
Delville, A. \j Langmuir, 1992, {\it 8}, 1796.
Curry, J.E.; Mcquarrie, D.A.
\j Langmuir, 1992, {\it 8}, 1026.
Takahashi, T.; Yamaguchi, M.
\j J. Colloid and Interface Sci., 1991, {\it 146}, 556.
Ramsay, J.D.F.; Swanton, S.W.; Bunce, J.
\j J. Chem. Soc. Faraday Trans., 1990, {\it 86}, 3919.
Delville, A.; Laszlo, P.
\j Langmuir, 1990, {\it 6}, 1289.}
\refdef{CR}{Kokufuta, E.; Zhang, Y.Q.; Tanaka, T.
\j J. Biomaterials Science-Polymer Edition, 1994, {\it 6}, 35.}
\refdef{Sekimoto}{Kuroki, Y.; Sekimoto, K.
\j Europhys. Lett., 1994, {\it 26}, 227.  }
\refdef{release}{Li, Y.X.; Kissel, T.
\j J. Controlled Release, 1993, {\it 27}, 247.  }
\refdef{Tanaka}{Matsuo, E.S.; Tanaka, T.
\j Nature, 1992, {\it 358}, 482.  }
\refdef{Kinetics}{Li, Y.; Tanaka, T.
\j J. Chem. Phys., 1990, {\it 92}, 1365.
Rossi, G.; Mazich, K.A. \j Phys. Rev. E, 1993, {\it 48}, 1182.
Rossi, G.; Mazich, K.A. \j Phys. Rev. A, 1991, {\it 44}, R4793.  }
\refdef{rel2}{Colombo, P.; Gazzaniga, A.; Caramella, C.; Conte, U.;
Manna, A.L.
\j Acta Pharm. Technol., 1987, {\it 33} (1), 15.
Conte, U.; Colombo, P.; Gazzaniga, A.; Sangalli, M.E.; Manna, A.L.
\j Biomaterials, 1988, {\it 9}, 489;
Peppas, N.A.; Wu, J.C.; van Meerwall, E.D.
\j Macromolecules, 1994, {\it 27}, 5626.
Brown, D.; Bae, Y.H.; Kim, S.W.
\j Macromolecules, 1994, {\it 27}, 4952.}
\refdef{LandauLifshitz}{Landau, L.D.; Lifshitz, E.M. In {\it Theory of
Elasticity}; Pergamon Press: New York, 1986.}
\refdef{qu}{We note that a cubic
term may play an important role for the compression of non-bridging brushes,
as suggested by the self-consistent field theory \cite{Mi}.
For triblock copolymer mesogels, however, the B regions consist
of bridging and non-bridging B chains.
Those B chains bridging different
A domains should have a step-function profile.
Therefore, for small deviations from equilibrium,
we expect that the leading term in the free energy is the quadratic term.}
\refdef{Mi}{Milner, S.T.; Witten, T.A.; Cates, M.E. \j Macromolecules, 1988,
{\it 21}, 2610.}

\begin{figure}
\caption{The swelling geometry for a single layer. The solvent
penetrates from the left and the gel is assumed infinite towards
the right and out of the plane of the page. The swelling profile
is characterized by $h(x)$.}
\label{one}
\end{figure}

\begin{figure}
\caption{The swelling geometry for multiple layers.
Note that in this case the swelling of the inner layers causes
a large bending at the outer layers.}
\label{two}
\end{figure}


\begin{references}
\reflist{}

\end{references}
\end{document}